\documentclass[%
 aip,
 amsmath,amssymb,
preprint,%
]{revtex4-1}

\usepackage{graphicx}
\usepackage{dcolumn}
\usepackage{bm}
\usepackage{hyperref}
\usepackage[utf8]{inputenc}
\usepackage[T1]{fontenc}
\usepackage{mathptmx}
\usepackage{etoolbox}

\makeatletter
\def\@email#1#2{%
 \endgroup
 \patchcmd{\titleblock@produce}
  {\frontmatter@RRAPformat}
  {\frontmatter@RRAPformat{\produce@RRAP{*#1\href{mailto:#2}{#2}}}\frontmatter@RRAPformat}
  {}{}
}%
\makeatother
\begin{document}


\title{Functionalized Tellurene; a candidate large-gap 2D Topological Insulator}
\author{Raghottam M. Sattigeri}
\author{Prafulla K. Jha}%
\email{prafullaj@yahoo.com}
\affiliation{$ ^{1} $Department of Physics, Faculty of Science, The Maharaja Sayajirao University of Baroda, Vadodara-390002, Gujarat, India}%
%


\begin{abstract}
The discovery of group IV and V elemental Xene's which exhibit topologically non-trivial characters natively in their honeycomb lattice structure (HLS) has led to extensive efforts in realising analogous behaviour in group VI elemental monolayers. Although; it was theoretically concluded that group VI elemental monolayers cannot exist as HLS but recent experimental evidence suggests otherwise. In this letter we report that, HLS of group VI elemental monolayer (such as, Tellurene) can be realised to be dynamically stable when functionzalised with Oxygen. The functionalization leads to, peculiar orbital filtering effects (OFE) and broken spatial inversion symmetry which gives rise to the non-trivial topological character. The exotic quantum behaviour of this system is characterized by, spin-orbit coupling induced large-gap $\approx$ 0.36 eV with isolated Dirac cone along the edges indicating perspective room temperature spin-transport applications. Further investigations of spin Hall conductivity and the Berry curvatures unravel high conductivity as compared to previously explored Xene's. The non-trivial topological character is quantified in terms of the $\mathbb{Z}_2$ invaraint as $\nu =$ 1 and Chern number $\mathit{C} =$ 1. Also, for practical purposes, we report that, \textit{h}BN/TeO/\textit{h}BN quantum-wells can be strain engineered to realize a sizable non-trivial gap ($\approx$ 0.11 eV). We finally conclude that, functionalization of group VI elemental monolayer with Oxygen gives rise to, exotic quantum properties which are robust against surface oxidation and degradations while providing viable electronic degrees of freedom for spintronic applications. 
\end{abstract}

\maketitle

%


Topological Insulators (TI) are exotic quantum phase of matter characterized by the \textit{D} dimensional bulk gap and the \textit{D$-$1} dimensional conducting states which are robust against weak deformations and external perturbations \cite{1,2,3,4,5}. In two-dimensional (2D) systems this effect gives rise to, gapless spin-polarized helical edge states with insulating non-magnetic bulk gap \cite{6}. Since such systems are governed by the time-reversal symmetry, the dissipationless spin currents are protected against backscattering from non-magnetic impurities leading to quantized Hall conductance in absence of magnetic field \cite{7,8}. This makes them highly desirable candidates for spintronic, valleytronic, electronic and quantum computing applications. 

\begin{figure*}
	\includegraphics[width=16cm]{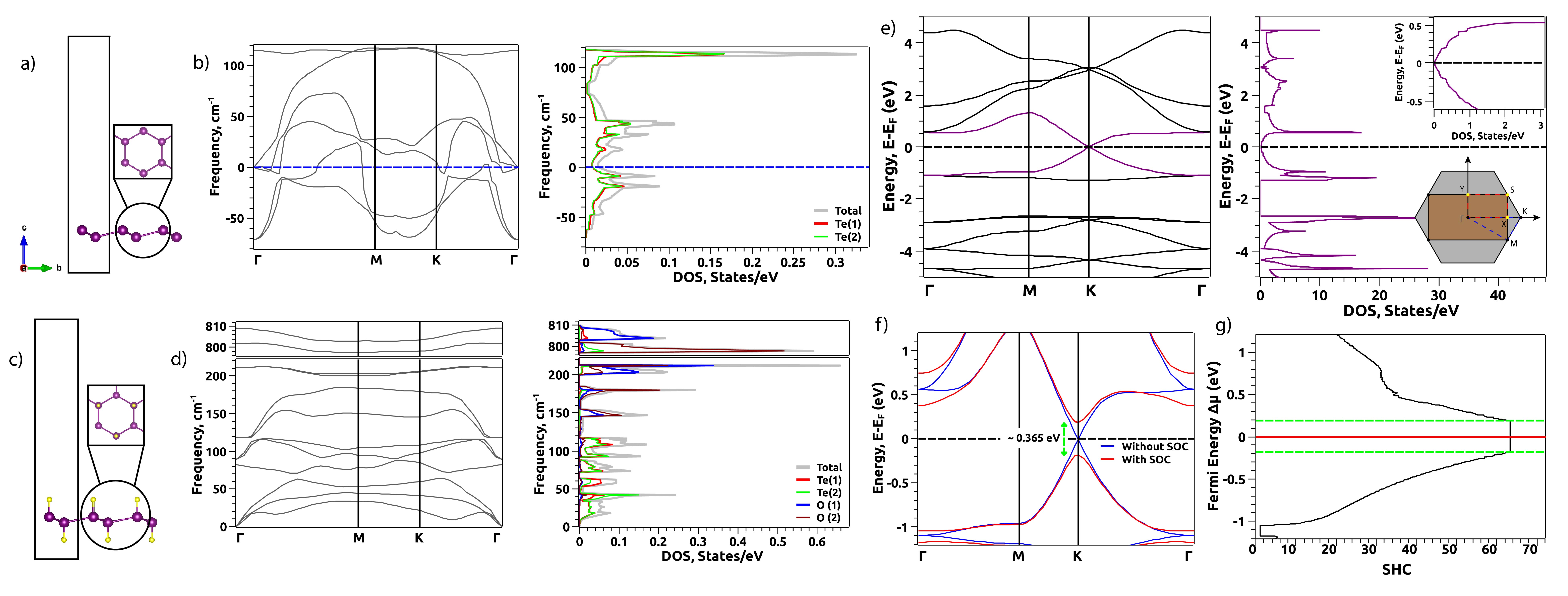}
	\caption{\label{fig1} a) HLS of monolayer Te. b) PDC alongside PhDOS of monolayer Te which exhibits imaginary modes indicating dynamically unstable structure. c) HLS of monolayer TeO wherein, Te monolayer is functionalized with Oxygen at alternative sublattice positions on either side of the monolayer. d) PDC alongside PhDOS of TeO exhibits absence of imaginary modes indicating dynamically stable structure. e) EBS alongside DOS indicates the semi-metallic nature of TeO (inset zoomed version DOS and the 2D BZ with highlighted edge BZ). f) EBS without and with SOC indicating non-trivial large-gap alongside the Fermi energy scan ($\Delta\mu$) of SHC ($\sigma^{\textit{spinz}}_{\textit{xy}} (\omega)$) in the units of $(\hbar/e)\Omega^{-1}cm^{-1}$ (red line indicates the Fermi energy and the green lines indicate the SOC induced band gap).}
\end{figure*}

Although, with the discovery of graphene several 2D TI based on group IV and V elements were theoretically predicted \cite{9,10,11,12,13,14,15,16,17,18,19,20} but, the phenomena was only experimentally realised and limited to HgTe/CdTe and InAs/GaSb quantum wells (QW) at ultra-low temperatures and ultra-high vacuum by the virtue of weak spin-orbit coupling (SOC) effects \cite{21}. However, this sparked several investigations based on \textit{first-principles} and theorical modelling methods which have driven the experimental efforts over the years. One of the major challenges is, realising large band gaps which would make room temperature applications of 2D TI materials viable. This has been well addressed by the functionalization of monolayers of group IV, V Xene's and MXene's etc., which leads to the -\textit{pz} orbital saturation and enhancement of the SOC by virtue of orbital filtering effects (OFE) \cite{7,13,22,23,24,25,26,27,28,29,30}. For example, it was experimentally shown that halogenation of Bi monolayers can host non-trivial TI states but, these states are quite sensitive due to their strong hybridizations with the substrates \cite{31}. However, the robustness of TI nature against surface oxidation and degradation is not well addressed which might hinder their practical applications.

Similar to group IV and V elemental monolayers, efforts were made to realise 2D TI nature in group VI Xene's with honeycomb lattice structure (HLS) based on Tellurium (Te) and Selenium (Se). But, the failure is attributed to electronic configuration of group VI elements and their bulk structures which inhibit fostering the hexagonal HLS rendering them to exist in ring, chain, square and rectangular lattice structures \cite{20,32,33,34}. It was eventually concluded that, elemental monolayers of Te and Se \textit{cannot} exist in HLS \cite{35}. However, recent, experimental evidence suggests otherwise, i.e., thickness-dependent transition from Au-Te surface alloy to HLS of Te (i.e., 3 $\times$ 3 honeycomb-like superstructures) can be realised by employing the self-organization mechanism of Te atoms on Au(111) surface (when exposed to high temperature vapor deposition of Te in ultra-high vacuum) at 0.5 monolayer coverage \cite{36}. 

Inspired by this, in this letter we explore alternative approach to realise dynamically stable HLS of Te and Se. For this purpose, we perform Oxygen and Sulphur \textit{functionalization} of Te and Se monolayers since; the pristine monolayers are unstable which was found to be in agreement with previous studies \cite{35}. Off these, only Oxygen functionalised Te and Se and Sulphur functionalised Se were found to be dynamically stable. However, by the virtue of OFE and large SOC (due to high atomic number of Te) Oxygen functionalised Te exhibits a large-gap 2D TI nature with viable room temperature applications by exploiting the robust Dirac dispersion of spin-polarized edge states. Hence, we restrict our discussions to non-trivial nature of Oxygen functionalised Te (while the other systems are presented in the supplementary material (SM)) which is also analysed in terms of the $\mathbb{Z}_2$ invariant and the Chern number ($\mathit{C}$).


We implement first-principles based Density Functional Theory in Quantum Espresso package \cite{37} to obtain the ground state energy of the system (with a vacuum of \textit{c} = 25.00 \AA{} along the [001] crystal direction to isolate the interactions between periodic images) under plane wave self consistent formulation and calculate the electronic band structures (EBS), density of states (DOS) and orbital/elemental projected density of states (PDOS). We use, \textit{scalar-relativistic} and norm conserving Martins-Troullier pseudopotentials for calculations without SOC and \textit{fully-relativistic} Projector augemented wave pseudopotentials (for calculations with SOC) under generalized gradient approximation with Perdew-Burke-Ernzerhof type of exchange-correlation functional in our calculations \cite{38,39}. The converged value of kinetic energy cutoff was 80 Ry and the corresponding uniform momentum (\textbf{\textit{k}}) Monkhorst-Pack grid was of 8 $\times$ 8 $\times$ 1 \cite{40}. The structures were fully relaxed with force convergence criteria of $<$ 10$^{-6}$ a.u. The dynamic stability of the system is investigated in terms of Phonon dispersion curves (PDC) and Phonon density of states (PhDOS) under the Density Functional Perturbation Theory \cite{41} regime with a \textbf{\textit{q}}-mesh of 5 $\times$ 5 $\times$ 1. We then project the DFT wave functions onto the maximally localized Wannier functions (with convergence of the spread function) by using Wannier90 code \cite{42} to compute the quantum transport properties and create the tight-binding hamiltonian as an input for the WannierTools code \cite{43}. We compute the spin Hall conductivity (SHC) ($\sigma^{\textit{spinz}}_{\textit{xy}} (\omega)$) using the Kubo-Greenwood formula \cite{44,45} within the independent-particle approximation, and the Berry curvature ($\Omega_\textit{z} (k)$) and \textit{k}-resolved spin Berry curvature ($\Omega^{\textit{spinz}}_{\textit{xy}} (k)$) to understand the topological properties and the quantum transport phenomena. For this purpose, a dense \textbf{\textit{k}}-mesh of 100 $ \times $ 100 $ \times $ 1 is used; since large contributions of spin Berry curvatures occur in minute regions of \textbf{\textit{k}} space which leads to slow convergence \cite{46}. We finally compute the $\mathbb{Z}_2$ invariant, Chern number ($\mathit{C}$) and edge states using the WannierTools code \cite{43}.

\begin{figure*}
	\includegraphics[width=16cm]{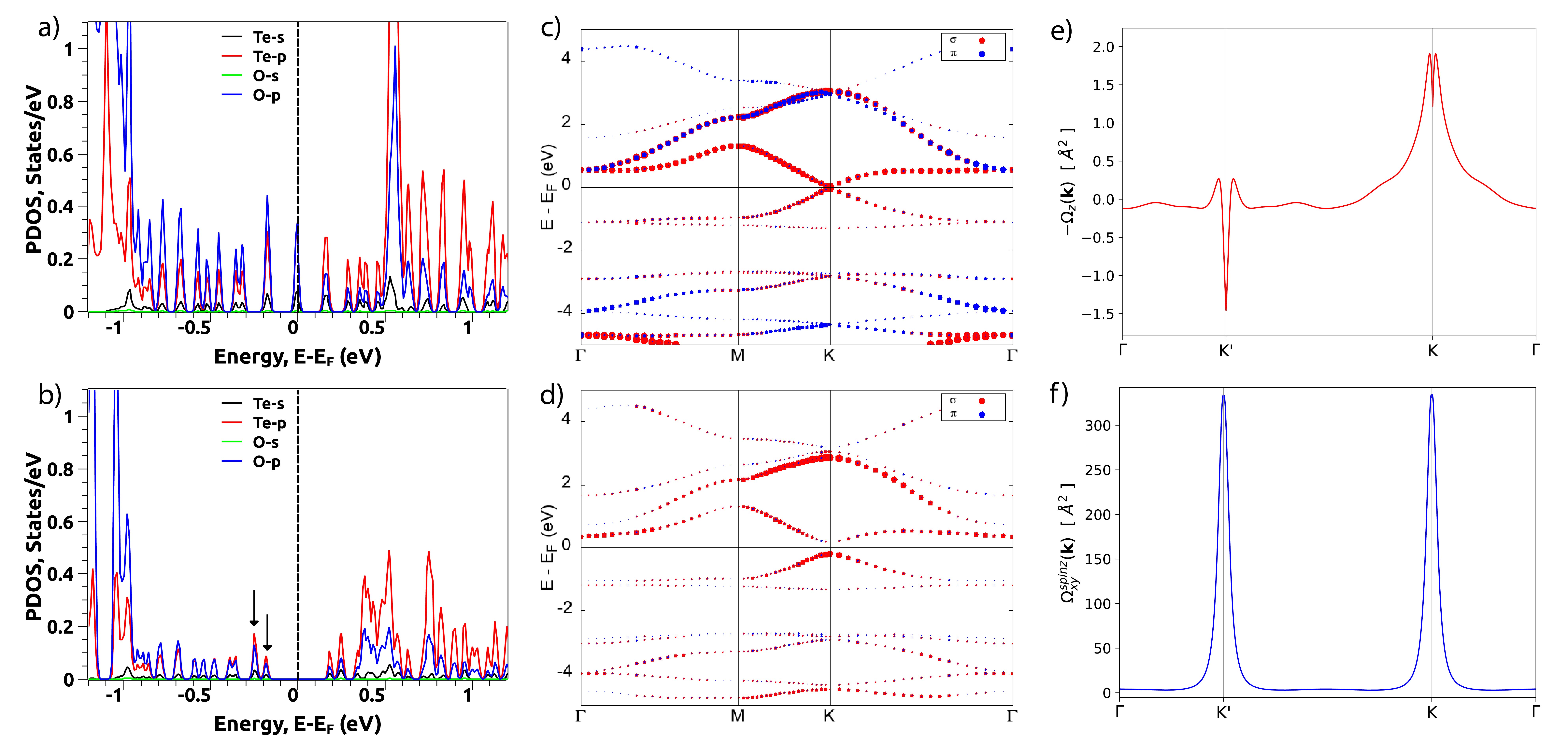}
	\caption{\label{fig2} PDOS of TeO (a) without and (b) with SOC indicating exchange of -\textit{p} orbitals across the Fermi highlighted by downward arrows. Orbital ($\pi$ and $\sigma$) projected EBS of TeO (c) without and (d) with SOC. e) Berry curvature ($\Omega_\textit{z} (k)$) mapped along the entire 2D BZ indicates opposite signs. f) \textit{k}-resolved spin Berry curvature ($\Omega^{\textit{spinz}}_{\textit{xy}} (k)$) mapped along the BZ valleys \textit{K'} and \textit{K}.}
\end{figure*}


Similar to graphene and several other group IV,V elemental monolayers we begin with designing of elemental monolayers composed of Te (and Se) with HLS as shown in Fig. \ref{fig1}(a) (and Fig. 1(b)(SM)). The optimized lattice constant of Te (and Se; presented in SM Table I) with HLS is \textit{a} = 4.94 \AA{}, with a buckling height of \textit{h} = 0.72 \AA{} and the angle between Te atoms $\approx$ 114.14$^{\circ}$. However, we find these structures to be dynamically unstable due to imaginary modes throughout the Brillouin Zone (BZ) as evident from Fig. \ref{fig1}(b) (and Fig. 1(a)(SM)), which is in agreement with previous studies \cite{35}. Such a HLS is not favoured due to, the electronic configuration of Te (and Se) which does not allow the formation of typical -\textit{sp} hybridizations and the reactive surface due to the dangling bonds \cite{47}. However, on functionalization with Oxygen at alternative sublattice positions on both sides of Te (and Se) monolayer (Fig. \ref{fig1}(c) and Fig. 1(c)(SM)) and Sulphur on Se (Fig. 1(f)(SM)), the imaginary modes are eliminated rendering the system (Tellurene Oxide-TeO, Selenene Oxide-SeO and Selenene Sulphide-SeS) to be dynamically stable (Fig. \ref{fig1}(d) and Fig. 1(d,e)(SM)) in the much sought-after HLS. This structure is governed by the \textit{D}$_{3d}$ point group symmetry which is lower than the \textit{D}$_{6h}$ point group symmetry of graphene; leading to broken spatial inversion symmetry in the system. The optimized lattice constant of TeO is, \textit{a} = 5.26 \AA{}, with reduced buckling height of \textit{h} = 0.57 \AA{} (which indicates viable structural design), layer thickness \textit{t} $\approx$ 4.18 \AA{}, $\angle$O-Te-Te $=$ 100.74$^{\circ}$ and the angle between Te atoms as $\approx$ 116.61$^{\circ}$. Apart from stability, the functionalization contributes by, saturating the out of plane -\textit{pz} orbitals (dangling bonds) of Te giving rise to the -\textit{sp}$^3$ hybridization and strong SOC. 

The 2D hexagonal BZ and the orthorhombic edge BZ are presented in inset of Fig. \ref{fig1}(e). It is evident that, when the SOC is not considered TeO exhibits graphene-like semi-metallic behaviour along the high symmetry point \textit{K} in the BZ hosting degenerate Dirac-cone states (Fig. \ref{fig1}(e)) (whereas, SeO and SeS are semi-conducting with gaps of the order of $\approx$ 10$^{-3}$ eV as evident from band structure presented in Fig. 2(SM)). The electronic band structure of TeO along the 2D hexagonal BZ path with and without SOC is presented in Fig. \ref{fig1}(f). SOC lifts off the degeneracy giving rise to a global large-gap of $\approx$ 0.365 eV due to strong SOC and OFE. The strong SOC is due to the domination of $\sigma$ orbitals around the Fermi, rather than the $\pi$ orbitals with weaker SOC as evident from the orbital projected band structures presented in Fig. \ref{fig2}(c-d) \cite{23} apart from the high atomic number of Te. Also, from the projected density of states (Fig. \ref{fig2}) it is evident that the orbital band inversion mechanism involves -\textit{p} orbitals of Te and O. This sort of band inversion along with large global gap indicates the potential non-trivial TI character of TeO. The -\textit{p} orbital near the vicinity of Fermi splits into two groups -\textit{pz} and -\textit{px}, -\textit{py} since the system is governed by \textit{D}$_{3d}$ point group symmetry and the \textit{D}$_3$ groups of the wave vector at Dirac points \textit{K} and \textit{K'} in the BZ. The flat bands (below the Fermi) made up of -\textit{px} and -\textit{py} orbitals and the massive Dirac cones make TeO unique as compared to graphene which indicates that TeO maybe a novel orbital analog of quantum anomalous Hall effect and Wigner crystallization \cite{19}. 

\begin{figure*}[ht]
	\includegraphics[width=16cm]{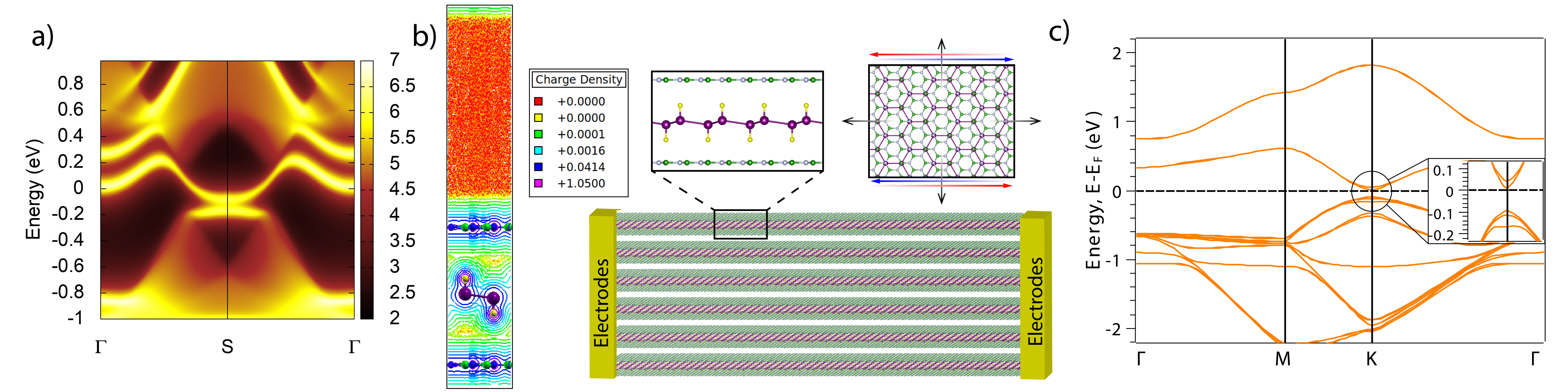}
	\caption{\label{fig3} a) Edge states along high symmetry point S in the edge BZ indicating towards bulk-boundary correspondence in TeO (temperature scale is in arbitrary units). b) Schematic of the van der Waals \textit{h}BN/TeO/\textit{h}BN quantum well heterostructure device subjected to uniform tensile strain. c) EBS of \textit{h}BN/TeO/\textit{h}BN indicating the persistence of non-trivial gap along high symmetry point K in the BZ.}
\end{figure*}

We further analyse the $\mathbb{Z}_2$ invariant by using Wilson loops method around Wannier Charge Centers \cite{43}; which turns out to be $\nu =$ 1, confirming the non-trivial TI character of TeO and the corresponding quantum spin Hall effect. The Berry curvature distribution in the entire BZ is presented in Fig. \ref{fig2}(e) for the n$^{th}$ occupied band. By integrating the Berry curvature throughout the BZ, we find the Chern number to be $ \mathit{C} =$ 1 which further confirms the non-trivial TI character of the system. Also, non-trivial TeO will host bulk-boundary correspondance which is a key ingredient in realising a potential 2D TI. For this purpose we compute the edge states along the high symmetry paths of orthorhombic edge BZ. From Fig. \ref{fig3}(a), it is clearly evident that, 2D TeO is a TI hosting robust conducting edge states along the high symmetry point S in the edge BZ. We further analyse the origin and nature of SHC in this system.

Since the strong SOC effects split the band degeneracies along the high symmetry point K in the hexagonal BZ with the Fermi level lying within the gap, we can expect significantly large SHC \cite{46}. This is because, the spin Berry curvature contributions (presented in Fig. \ref{fig2}(f)) from the positive and negative spins around Fermi do not cancel out excluding one sign of spin Berry curvature. The corresponding spin Berry curvature governed topology of TeO gives rise to unconventionally large transverse SHC ($\sigma^\textit{z}_{\textit{xy}}$) of the order of $\approx$ 63.25 $(\hbar/e)\Omega^{-1}cm^{-1}$ (evident from Fig. \ref{fig1}(g) which shows the Fermi energy scan $\delta\mu$ to study the SHC). This is superior as compared to other systems \cite{48,49} which indicates higher charge-spin-current conversion efficiency. The absence of quantization in the SHC can be attributd to the presence of buckling in the system \cite{48}. 

From the Berry curvature plot presented in Fig. \ref{fig2}(e) we observe sharp peaks in the valley region with opposite signs for \textit{K} and \textit{K'} which implies the possibility of valley Hall effects in the system. This implies that, when the TeO monolayer is subjected to voltage bias, the electrons from different valleys would experience opposite Lorentz forces leading to electronic motion in opposite directions perpendicular to the drift current thus creating charge-spin segregation making the system a potential quantum valley Hall insulator \cite{50}.

For practical purposes, apart from the free standing monolayers of TeO, we explore their quantum well heterostructures with $2 \times 2$ hexagonal-Boron Nitride (\textit{h}BN) presented in Fig. \ref{fig3}(b). We can sandwitch TeO between \textit{h}BN creating quantum wells of the form \textit{h}BN/TeO/\textit{h}BN with small lattice mismatch of $\approx$ 4.79\% and interlayer separation was varied from 3 \AA{} to 6 \AA{}. Upon optimization of this system under van der Waals appromixation \cite{51}, we observe that, the TeO and \textit{h}BN retain their original structures. When subjected to tensile strain as high as $\approx$ 14\% the system hosts a non-trivial gap of the order of $\approx$ 0.11 eV (evident from Fig. \ref{fig3}c) which is considerably high as compared to room temperature thermal energy and majorly arising from TeO states around the Fermi due to weak coupling with \textit{h}BN with interlayer separation of 6 \AA{}. This would electrically insulate the layers of TeO protecting the spin-polarised helical edge states. Such aggrangements are known to increase the number of edge transport channels to support the dissipationless charge-spin transport \cite{25}.

In summary, using first-principles based DFT calculations, we report a novel large-gap 2D TI phase in HLS of TeO (with a possibility of exploring SeO and SeS for their potential TI properties under modest strain/electric field). This non-trivial phase is characterised in terms of the $\mathbb{Z}_2$ invariant, Chern number $\mathit{C}$ and the edge state spectrum. From practical applications point of view, we propose, strained quantum well heterostructure of the form hBN/TeO/hBN which can be used to realise charge-spin transport phenomena at room temperature. We also discussed the origin of unconventionally large SHC in this system and the possibility of valley Hall insulating phenomena by analysing the Berry curvature and spin Berry curvature. Our predictions may further inspire experimental efforts to realise this novel phase of TeO with room temperature applications as spin Hall insulators / valley Hall insulators.

See the supplementary material for additional information on the structural parameters of Se, SeO, SeS and TeS along with their Phonon Dispersion Curves and the electronic band structures of SeO and SeS.

\section*{ACKNOWLEDGEMENT}
The authors acknowledge, the Interdisciplinary Centre for Mathematical and Computational Modelling UW (ICM UW), Warsaw, Poland for providing computational facilities. RMS acknowledges Dr. Som Narayan, Warsaw University of Technology, Poland and Dr. Trupti Gajaria, GSFC University, India for fruitful discussions and sugesstions.on

\subsection*{CONFLICT OF INTEREST}
The authors have no conflicts of interest to disclose.

\section*{AUTHORS’ CONTRIBUTIONS}
R.M.S. and P.K.J. contributed equally to this work.

\section*{DATA AVAILABILITY}
The data that support the findings of this work are available from
the corresponding author upon reasonable request.

\nocite{*}
\bibliography{aipsamp}

\providecommand{\noopsort}[1]{}\providecommand{\singleletter}[1]{#1}%
\begin{thebibliography}{51}%
\makeatletter
\providecommand \@ifxundefined [1]{%
 \@ifx{#1\undefined}
}%
\providecommand \@ifnum [1]{%
 \ifnum #1\expandafter \@firstoftwo
 \else \expandafter \@secondoftwo
 \fi
}%
\providecommand \@ifx [1]{%
 \ifx #1\expandafter \@firstoftwo
 \else \expandafter \@secondoftwo
 \fi
}%
\providecommand \natexlab [1]{#1}%
\providecommand \enquote  [1]{``#1''}%
\providecommand \bibnamefont  [1]{#1}%
\providecommand \bibfnamefont [1]{#1}%
\providecommand \citenamefont [1]{#1}%
\providecommand \href@noop [0]{\@secondoftwo}%
\providecommand \href [0]{\begingroup \@sanitize@url \@href}%
\providecommand \@href[1]{\@@startlink{#1}\@@href}%
\providecommand \@@href[1]{\endgroup#1\@@endlink}%
\providecommand \@sanitize@url [0]{\catcode `\\12\catcode `\$12\catcode
  `\&12\catcode `\#12\catcode `\^12\catcode `\_12\catcode `\%12\relax}%
\providecommand \@@startlink[1]{}%
\providecommand \@@endlink[0]{}%
\providecommand \url  [0]{\begingroup\@sanitize@url \@url }%
\providecommand \@url [1]{\endgroup\@href {#1}{\urlprefix }}%
\providecommand \urlprefix  [0]{URL }%
\providecommand \Eprint [0]{\href }%
\providecommand \doibase [0]{http://dx.doi.org/}%
\providecommand \selectlanguage [0]{\@gobble}%
\providecommand \bibinfo  [0]{\@secondoftwo}%
\providecommand \bibfield  [0]{\@secondoftwo}%
\providecommand \translation [1]{[#1]}%
\providecommand \BibitemOpen [0]{}%
\providecommand \bibitemStop [0]{}%
\providecommand \bibitemNoStop [0]{.\EOS\space}%
\providecommand \EOS [0]{\spacefactor3000\relax}%
\providecommand \BibitemShut  [1]{\csname bibitem#1\endcsname}%
\let\auto@bib@innerbib\@empty
\bibitem [{\citenamefont {Kane}\ and\ \citenamefont
  {Mele}(2005{\natexlab{a}})}]{1}%
  \BibitemOpen
  \bibfield  {author} {\bibinfo {author} {\bibfnamefont {C.~L.}\ \bibnamefont
  {Kane}}\ and\ \bibinfo {author} {\bibfnamefont {E.~J.}\ \bibnamefont
  {Mele}},\ }\href@noop {} {\bibfield  {journal} {\bibinfo  {journal} {Physical
  review letters}\ }\textbf {\bibinfo {volume} {95}},\ \bibinfo {pages}
  {146802} (\bibinfo {year} {2005}{\natexlab{a}})}\BibitemShut {NoStop}%
\bibitem [{\citenamefont {Kane}\ and\ \citenamefont
  {Mele}(2005{\natexlab{b}})}]{2}%
  \BibitemOpen
  \bibfield  {author} {\bibinfo {author} {\bibfnamefont {C.~L.}\ \bibnamefont
  {Kane}}\ and\ \bibinfo {author} {\bibfnamefont {E.~J.}\ \bibnamefont
  {Mele}},\ }\href@noop {} {\bibfield  {journal} {\bibinfo  {journal} {Physical
  review letters}\ }\textbf {\bibinfo {volume} {95}},\ \bibinfo {pages}
  {226801} (\bibinfo {year} {2005}{\natexlab{b}})}\BibitemShut {NoStop}%
\bibitem [{\citenamefont {Bernevig}\ and\ \citenamefont {Zhang}(2006)}]{3}%
  \BibitemOpen
  \bibfield  {author} {\bibinfo {author} {\bibfnamefont {B.~A.}\ \bibnamefont
  {Bernevig}}\ and\ \bibinfo {author} {\bibfnamefont {S.-C.}\ \bibnamefont
  {Zhang}},\ }\href@noop {} {\bibfield  {journal} {\bibinfo  {journal}
  {Physical review letters}\ }\textbf {\bibinfo {volume} {96}},\ \bibinfo
  {pages} {106802} (\bibinfo {year} {2006})}\BibitemShut {NoStop}%
\bibitem [{\citenamefont {Hasan}\ and\ \citenamefont {Kane}(2010)}]{4}%
  \BibitemOpen
  \bibfield  {author} {\bibinfo {author} {\bibfnamefont {M.~Z.}\ \bibnamefont
  {Hasan}}\ and\ \bibinfo {author} {\bibfnamefont {C.~L.}\ \bibnamefont
  {Kane}},\ }\href@noop {} {\bibfield  {journal} {\bibinfo  {journal} {Reviews
  of modern physics}\ }\textbf {\bibinfo {volume} {82}},\ \bibinfo {pages}
  {3045} (\bibinfo {year} {2010})}\BibitemShut {NoStop}%
\bibitem [{\citenamefont {Han}, \citenamefont {Kang},\ and\ \citenamefont
  {Jeon}(2020)}]{5}%
  \BibitemOpen
  \bibfield  {author} {\bibinfo {author} {\bibfnamefont {C.}~\bibnamefont
  {Han}}, \bibinfo {author} {\bibfnamefont {M.}~\bibnamefont {Kang}}, \ and\
  \bibinfo {author} {\bibfnamefont {H.}~\bibnamefont {Jeon}},\ }\href@noop {}
  {\bibfield  {journal} {\bibinfo  {journal} {ACS Photonics}\ }\textbf
  {\bibinfo {volume} {7}},\ \bibinfo {pages} {2027--2036} (\bibinfo {year}
  {2020})}\BibitemShut {NoStop}%
\bibitem [{\citenamefont {Chen}\ \emph {et~al.}(2018)\citenamefont {Chen},
  \citenamefont {Pai}, \citenamefont {Chan}, \citenamefont {Sun}, \citenamefont
  {Xu}, \citenamefont {Lin}, \citenamefont {Chou}, \citenamefont {Fedorov},\
  and\ \citenamefont {Chiang}}]{6}%
  \BibitemOpen
  \bibfield  {author} {\bibinfo {author} {\bibfnamefont {P.}~\bibnamefont
  {Chen}}, \bibinfo {author} {\bibfnamefont {W.~W.}\ \bibnamefont {Pai}},
  \bibinfo {author} {\bibfnamefont {Y.-H.}\ \bibnamefont {Chan}}, \bibinfo
  {author} {\bibfnamefont {W.-L.}\ \bibnamefont {Sun}}, \bibinfo {author}
  {\bibfnamefont {C.-Z.}\ \bibnamefont {Xu}}, \bibinfo {author} {\bibfnamefont
  {D.-S.}\ \bibnamefont {Lin}}, \bibinfo {author} {\bibfnamefont
  {M.}~\bibnamefont {Chou}}, \bibinfo {author} {\bibfnamefont {A.-V.}\
  \bibnamefont {Fedorov}}, \ and\ \bibinfo {author} {\bibfnamefont {T.-C.}\
  \bibnamefont {Chiang}},\ }\href@noop {} {\bibfield  {journal} {\bibinfo
  {journal} {Nature communications}\ }\textbf {\bibinfo {volume} {9}},\
  \bibinfo {pages} {1--7} (\bibinfo {year} {2018})}\BibitemShut {NoStop}%
\bibitem [{\citenamefont {Hu}\ \emph {et~al.}(2018)\citenamefont {Hu},
  \citenamefont {Lyu}, \citenamefont {Zhang}, \citenamefont {Wang},
  \citenamefont {Ji},\ and\ \citenamefont {Li}}]{7}%
  \BibitemOpen
  \bibfield  {author} {\bibinfo {author} {\bibfnamefont {X.-K.}\ \bibnamefont
  {Hu}}, \bibinfo {author} {\bibfnamefont {J.-K.}\ \bibnamefont {Lyu}},
  \bibinfo {author} {\bibfnamefont {C.-W.}\ \bibnamefont {Zhang}}, \bibinfo
  {author} {\bibfnamefont {P.-J.}\ \bibnamefont {Wang}}, \bibinfo {author}
  {\bibfnamefont {W.-X.}\ \bibnamefont {Ji}}, \ and\ \bibinfo {author}
  {\bibfnamefont {P.}~\bibnamefont {Li}},\ }\href@noop {} {\bibfield  {journal}
  {\bibinfo  {journal} {Physical Chemistry Chemical Physics}\ }\textbf
  {\bibinfo {volume} {20}},\ \bibinfo {pages} {13632--13636} (\bibinfo {year}
  {2018})}\BibitemShut {NoStop}%
\bibitem [{\citenamefont {Gusev}\ \emph {et~al.}(2020)\citenamefont {Gusev},
  \citenamefont {Olshanetsky}, \citenamefont {Hernandez}, \citenamefont
  {Raichev}, \citenamefont {Mikhailov},\ and\ \citenamefont {Dvoretsky}}]{8}%
  \BibitemOpen
  \bibfield  {author} {\bibinfo {author} {\bibfnamefont {G.}~\bibnamefont
  {Gusev}}, \bibinfo {author} {\bibfnamefont {E.}~\bibnamefont {Olshanetsky}},
  \bibinfo {author} {\bibfnamefont {F.}~\bibnamefont {Hernandez}}, \bibinfo
  {author} {\bibfnamefont {O.}~\bibnamefont {Raichev}}, \bibinfo {author}
  {\bibfnamefont {N.}~\bibnamefont {Mikhailov}}, \ and\ \bibinfo {author}
  {\bibfnamefont {S.}~\bibnamefont {Dvoretsky}},\ }\href@noop {} {\bibfield
  {journal} {\bibinfo  {journal} {Physical Review B}\ }\textbf {\bibinfo
  {volume} {101}},\ \bibinfo {pages} {241302} (\bibinfo {year}
  {2020})}\BibitemShut {NoStop}%
\bibitem [{\citenamefont {Yao}\ \emph {et~al.}(2007)\citenamefont {Yao},
  \citenamefont {Ye}, \citenamefont {Qi}, \citenamefont {Zhang},\ and\
  \citenamefont {Fang}}]{9}%
  \BibitemOpen
  \bibfield  {author} {\bibinfo {author} {\bibfnamefont {Y.}~\bibnamefont
  {Yao}}, \bibinfo {author} {\bibfnamefont {F.}~\bibnamefont {Ye}}, \bibinfo
  {author} {\bibfnamefont {X.-L.}\ \bibnamefont {Qi}}, \bibinfo {author}
  {\bibfnamefont {S.-C.}\ \bibnamefont {Zhang}}, \ and\ \bibinfo {author}
  {\bibfnamefont {Z.}~\bibnamefont {Fang}},\ }\href@noop {} {\bibfield
  {journal} {\bibinfo  {journal} {Physical Review B}\ }\textbf {\bibinfo
  {volume} {75}},\ \bibinfo {pages} {041401} (\bibinfo {year}
  {2007})}\BibitemShut {NoStop}%
\bibitem [{\citenamefont {Konschuh}, \citenamefont {Gmitra},\ and\
  \citenamefont {Fabian}(2010)}]{10}%
  \BibitemOpen
  \bibfield  {author} {\bibinfo {author} {\bibfnamefont {S.}~\bibnamefont
  {Konschuh}}, \bibinfo {author} {\bibfnamefont {M.}~\bibnamefont {Gmitra}}, \
  and\ \bibinfo {author} {\bibfnamefont {J.}~\bibnamefont {Fabian}},\
  }\href@noop {} {\bibfield  {journal} {\bibinfo  {journal} {Physical Review
  B}\ }\textbf {\bibinfo {volume} {82}},\ \bibinfo {pages} {245412} (\bibinfo
  {year} {2010})}\BibitemShut {NoStop}%
\bibitem [{\citenamefont {Deng}\ \emph {et~al.}(2018)\citenamefont {Deng},
  \citenamefont {Xia}, \citenamefont {Ma}, \citenamefont {Chen}, \citenamefont
  {Shan}, \citenamefont {Zhai}, \citenamefont {Li}, \citenamefont {Zhao},
  \citenamefont {Xu}, \citenamefont {Duan} \emph {et~al.}}]{11}%
  \BibitemOpen
  \bibfield  {author} {\bibinfo {author} {\bibfnamefont {J.}~\bibnamefont
  {Deng}}, \bibinfo {author} {\bibfnamefont {B.}~\bibnamefont {Xia}}, \bibinfo
  {author} {\bibfnamefont {X.}~\bibnamefont {Ma}}, \bibinfo {author}
  {\bibfnamefont {H.}~\bibnamefont {Chen}}, \bibinfo {author} {\bibfnamefont
  {H.}~\bibnamefont {Shan}}, \bibinfo {author} {\bibfnamefont {X.}~\bibnamefont
  {Zhai}}, \bibinfo {author} {\bibfnamefont {B.}~\bibnamefont {Li}}, \bibinfo
  {author} {\bibfnamefont {A.}~\bibnamefont {Zhao}}, \bibinfo {author}
  {\bibfnamefont {Y.}~\bibnamefont {Xu}}, \bibinfo {author} {\bibfnamefont
  {W.}~\bibnamefont {Duan}},  \emph {et~al.},\ }\href@noop {} {\bibfield
  {journal} {\bibinfo  {journal} {Nature materials}\ }\textbf {\bibinfo
  {volume} {17}},\ \bibinfo {pages} {1081--1086} (\bibinfo {year}
  {2018})}\BibitemShut {NoStop}%
\bibitem [{\citenamefont {Liu}, \citenamefont {Feng},\ and\ \citenamefont
  {Yao}(2011)}]{12}%
  \BibitemOpen
  \bibfield  {author} {\bibinfo {author} {\bibfnamefont {C.-C.}\ \bibnamefont
  {Liu}}, \bibinfo {author} {\bibfnamefont {W.}~\bibnamefont {Feng}}, \ and\
  \bibinfo {author} {\bibfnamefont {Y.}~\bibnamefont {Yao}},\ }\href@noop {}
  {\bibfield  {journal} {\bibinfo  {journal} {Physical review letters}\
  }\textbf {\bibinfo {volume} {107}},\ \bibinfo {pages} {076802} (\bibinfo
  {year} {2011})}\BibitemShut {NoStop}%
\bibitem [{\citenamefont {Xu}\ \emph {et~al.}(2013)\citenamefont {Xu},
  \citenamefont {Yan}, \citenamefont {Zhang}, \citenamefont {Wang},
  \citenamefont {Xu}, \citenamefont {Tang}, \citenamefont {Duan},\ and\
  \citenamefont {Zhang}}]{13}%
  \BibitemOpen
  \bibfield  {author} {\bibinfo {author} {\bibfnamefont {Y.}~\bibnamefont
  {Xu}}, \bibinfo {author} {\bibfnamefont {B.}~\bibnamefont {Yan}}, \bibinfo
  {author} {\bibfnamefont {H.-J.}\ \bibnamefont {Zhang}}, \bibinfo {author}
  {\bibfnamefont {J.}~\bibnamefont {Wang}}, \bibinfo {author} {\bibfnamefont
  {G.}~\bibnamefont {Xu}}, \bibinfo {author} {\bibfnamefont {P.}~\bibnamefont
  {Tang}}, \bibinfo {author} {\bibfnamefont {W.}~\bibnamefont {Duan}}, \ and\
  \bibinfo {author} {\bibfnamefont {S.-C.}\ \bibnamefont {Zhang}},\ }\href@noop
  {} {\bibfield  {journal} {\bibinfo  {journal} {Physical review letters}\
  }\textbf {\bibinfo {volume} {111}},\ \bibinfo {pages} {136804} (\bibinfo
  {year} {2013})}\BibitemShut {NoStop}%
\bibitem [{\citenamefont {Liu}\ \emph {et~al.}(2015)\citenamefont {Liu},
  \citenamefont {Zhang}, \citenamefont {Abdalla}, \citenamefont {Fazzio},\ and\
  \citenamefont {Zunger}}]{14}%
  \BibitemOpen
  \bibfield  {author} {\bibinfo {author} {\bibfnamefont {Q.}~\bibnamefont
  {Liu}}, \bibinfo {author} {\bibfnamefont {X.}~\bibnamefont {Zhang}}, \bibinfo
  {author} {\bibfnamefont {L.}~\bibnamefont {Abdalla}}, \bibinfo {author}
  {\bibfnamefont {A.}~\bibnamefont {Fazzio}}, \ and\ \bibinfo {author}
  {\bibfnamefont {A.}~\bibnamefont {Zunger}},\ }\href@noop {} {\bibfield
  {journal} {\bibinfo  {journal} {Nano letters}\ }\textbf {\bibinfo {volume}
  {15}},\ \bibinfo {pages} {1222--1228} (\bibinfo {year} {2015})}\BibitemShut
  {NoStop}%
\bibitem [{\citenamefont {Yang}\ \emph {et~al.}(2017)\citenamefont {Yang},
  \citenamefont {Xu}, \citenamefont {Liu}, \citenamefont {Jin}, \citenamefont
  {Zhang},\ and\ \citenamefont {Ding}}]{15}%
  \BibitemOpen
  \bibfield  {author} {\bibinfo {author} {\bibfnamefont {G.}~\bibnamefont
  {Yang}}, \bibinfo {author} {\bibfnamefont {Z.}~\bibnamefont {Xu}}, \bibinfo
  {author} {\bibfnamefont {Z.}~\bibnamefont {Liu}}, \bibinfo {author}
  {\bibfnamefont {S.}~\bibnamefont {Jin}}, \bibinfo {author} {\bibfnamefont
  {H.}~\bibnamefont {Zhang}}, \ and\ \bibinfo {author} {\bibfnamefont
  {Z.}~\bibnamefont {Ding}},\ }\href@noop {} {\bibfield  {journal} {\bibinfo
  {journal} {The Journal of Physical Chemistry C}\ }\textbf {\bibinfo {volume}
  {121}},\ \bibinfo {pages} {12945--12952} (\bibinfo {year}
  {2017})}\BibitemShut {NoStop}%
\bibitem [{\citenamefont {Wang}\ \emph
  {et~al.}(2016{\natexlab{a}})\citenamefont {Wang}, \citenamefont {Zhang},
  \citenamefont {Ji}, \citenamefont {Zhang}, \citenamefont {Li}, \citenamefont
  {Wang}, \citenamefont {Ren}, \citenamefont {Chen},\ and\ \citenamefont
  {Yuan}}]{16}%
  \BibitemOpen
  \bibfield  {author} {\bibinfo {author} {\bibfnamefont {Y.-p.}\ \bibnamefont
  {Wang}}, \bibinfo {author} {\bibfnamefont {C.-w.}\ \bibnamefont {Zhang}},
  \bibinfo {author} {\bibfnamefont {W.-x.}\ \bibnamefont {Ji}}, \bibinfo
  {author} {\bibfnamefont {R.-w.}\ \bibnamefont {Zhang}}, \bibinfo {author}
  {\bibfnamefont {P.}~\bibnamefont {Li}}, \bibinfo {author} {\bibfnamefont
  {P.-j.}\ \bibnamefont {Wang}}, \bibinfo {author} {\bibfnamefont {M.-j.}\
  \bibnamefont {Ren}}, \bibinfo {author} {\bibfnamefont {X.-l.}\ \bibnamefont
  {Chen}}, \ and\ \bibinfo {author} {\bibfnamefont {M.}~\bibnamefont {Yuan}},\
  }\href@noop {} {\bibfield  {journal} {\bibinfo  {journal} {Journal of Physics
  D: Applied Physics}\ }\textbf {\bibinfo {volume} {49}},\ \bibinfo {pages}
  {055305} (\bibinfo {year} {2016}{\natexlab{a}})}\BibitemShut {NoStop}%
\bibitem [{\citenamefont {Wang}\ \emph
  {et~al.}(2016{\natexlab{b}})\citenamefont {Wang}, \citenamefont {Ji},
  \citenamefont {Zhang}, \citenamefont {Li}, \citenamefont {Li}, \citenamefont
  {Ren}, \citenamefont {Chen}, \citenamefont {Yuan},\ and\ \citenamefont
  {Wang}}]{17}%
  \BibitemOpen
  \bibfield  {author} {\bibinfo {author} {\bibfnamefont {Y.-p.}\ \bibnamefont
  {Wang}}, \bibinfo {author} {\bibfnamefont {W.-x.}\ \bibnamefont {Ji}},
  \bibinfo {author} {\bibfnamefont {C.-w.}\ \bibnamefont {Zhang}}, \bibinfo
  {author} {\bibfnamefont {P.}~\bibnamefont {Li}}, \bibinfo {author}
  {\bibfnamefont {F.}~\bibnamefont {Li}}, \bibinfo {author} {\bibfnamefont
  {M.-j.}\ \bibnamefont {Ren}}, \bibinfo {author} {\bibfnamefont {X.-L.}\
  \bibnamefont {Chen}}, \bibinfo {author} {\bibfnamefont {M.}~\bibnamefont
  {Yuan}}, \ and\ \bibinfo {author} {\bibfnamefont {P.-j.}\ \bibnamefont
  {Wang}},\ }\href@noop {} {\bibfield  {journal} {\bibinfo  {journal}
  {Scientific reports}\ }\textbf {\bibinfo {volume} {6}},\ \bibinfo {pages}
  {1--8} (\bibinfo {year} {2016}{\natexlab{b}})}\BibitemShut {NoStop}%
\bibitem [{\citenamefont {Zhao}, \citenamefont {Zhang},\ and\ \citenamefont
  {Li}(2015)}]{18}%
  \BibitemOpen
  \bibfield  {author} {\bibinfo {author} {\bibfnamefont {M.}~\bibnamefont
  {Zhao}}, \bibinfo {author} {\bibfnamefont {X.}~\bibnamefont {Zhang}}, \ and\
  \bibinfo {author} {\bibfnamefont {L.}~\bibnamefont {Li}},\ }\href@noop {}
  {\bibfield  {journal} {\bibinfo  {journal} {Scientific reports}\ }\textbf
  {\bibinfo {volume} {5}},\ \bibinfo {pages} {1--7} (\bibinfo {year}
  {2015})}\BibitemShut {NoStop}%
\bibitem [{\citenamefont {Song}\ \emph {et~al.}(2014)\citenamefont {Song},
  \citenamefont {Liu}, \citenamefont {Yang}, \citenamefont {Han}, \citenamefont
  {Ye}, \citenamefont {Fu}, \citenamefont {Yang}, \citenamefont {Niu},
  \citenamefont {Lu},\ and\ \citenamefont {Yao}}]{19}%
  \BibitemOpen
  \bibfield  {author} {\bibinfo {author} {\bibfnamefont {Z.}~\bibnamefont
  {Song}}, \bibinfo {author} {\bibfnamefont {C.-C.}\ \bibnamefont {Liu}},
  \bibinfo {author} {\bibfnamefont {J.}~\bibnamefont {Yang}}, \bibinfo {author}
  {\bibfnamefont {J.}~\bibnamefont {Han}}, \bibinfo {author} {\bibfnamefont
  {M.}~\bibnamefont {Ye}}, \bibinfo {author} {\bibfnamefont {B.}~\bibnamefont
  {Fu}}, \bibinfo {author} {\bibfnamefont {Y.}~\bibnamefont {Yang}}, \bibinfo
  {author} {\bibfnamefont {Q.}~\bibnamefont {Niu}}, \bibinfo {author}
  {\bibfnamefont {J.}~\bibnamefont {Lu}}, \ and\ \bibinfo {author}
  {\bibfnamefont {Y.}~\bibnamefont {Yao}},\ }\href@noop {} {\bibfield
  {journal} {\bibinfo  {journal} {NPG Asia Materials}\ }\textbf {\bibinfo
  {volume} {6}},\ \bibinfo {pages} {e147--e147} (\bibinfo {year}
  {2014})}\BibitemShut {NoStop}%
\bibitem [{\citenamefont {Xian}\ \emph {et~al.}(2017)\citenamefont {Xian},
  \citenamefont {Paz}, \citenamefont {Bianco}, \citenamefont {Ajayan},\ and\
  \citenamefont {Rubio}}]{20}%
  \BibitemOpen
  \bibfield  {author} {\bibinfo {author} {\bibfnamefont {L.}~\bibnamefont
  {Xian}}, \bibinfo {author} {\bibfnamefont {A.~P.}\ \bibnamefont {Paz}},
  \bibinfo {author} {\bibfnamefont {E.}~\bibnamefont {Bianco}}, \bibinfo
  {author} {\bibfnamefont {P.~M.}\ \bibnamefont {Ajayan}}, \ and\ \bibinfo
  {author} {\bibfnamefont {A.}~\bibnamefont {Rubio}},\ }\href@noop {}
  {\bibfield  {journal} {\bibinfo  {journal} {2D Materials}\ }\textbf {\bibinfo
  {volume} {4}},\ \bibinfo {pages} {041003} (\bibinfo {year}
  {2017})}\BibitemShut {NoStop}%
\bibitem [{\citenamefont {Zhang}\ \emph {et~al.}(2017)\citenamefont {Zhang},
  \citenamefont {Zhang}, \citenamefont {Ji}, \citenamefont {Yan},\ and\
  \citenamefont {Yao}}]{21}%
  \BibitemOpen
  \bibfield  {author} {\bibinfo {author} {\bibfnamefont {R.-W.}\ \bibnamefont
  {Zhang}}, \bibinfo {author} {\bibfnamefont {C.-W.}\ \bibnamefont {Zhang}},
  \bibinfo {author} {\bibfnamefont {W.-X.}\ \bibnamefont {Ji}}, \bibinfo
  {author} {\bibfnamefont {S.-S.}\ \bibnamefont {Yan}}, \ and\ \bibinfo
  {author} {\bibfnamefont {Y.-G.}\ \bibnamefont {Yao}},\ }\href@noop {}
  {\bibfield  {journal} {\bibinfo  {journal} {Nanoscale}\ }\textbf {\bibinfo
  {volume} {9}},\ \bibinfo {pages} {8207--8212} (\bibinfo {year}
  {2017})}\BibitemShut {NoStop}%
\bibitem [{\citenamefont {Zhou}\ \emph {et~al.}(2014)\citenamefont {Zhou},
  \citenamefont {Ming}, \citenamefont {Liu}, \citenamefont {Wang},
  \citenamefont {Li},\ and\ \citenamefont {Liu}}]{22}%
  \BibitemOpen
  \bibfield  {author} {\bibinfo {author} {\bibfnamefont {M.}~\bibnamefont
  {Zhou}}, \bibinfo {author} {\bibfnamefont {W.}~\bibnamefont {Ming}}, \bibinfo
  {author} {\bibfnamefont {Z.}~\bibnamefont {Liu}}, \bibinfo {author}
  {\bibfnamefont {Z.}~\bibnamefont {Wang}}, \bibinfo {author} {\bibfnamefont
  {P.}~\bibnamefont {Li}}, \ and\ \bibinfo {author} {\bibfnamefont
  {F.}~\bibnamefont {Liu}},\ }\href@noop {} {\bibfield  {journal} {\bibinfo
  {journal} {Proceedings of the National Academy of Sciences}\ }\textbf
  {\bibinfo {volume} {111}},\ \bibinfo {pages} {14378--14381} (\bibinfo {year}
  {2014})}\BibitemShut {NoStop}%
\bibitem [{\citenamefont {Si}\ \emph {et~al.}(2014)\citenamefont {Si},
  \citenamefont {Liu}, \citenamefont {Xu}, \citenamefont {Wu}, \citenamefont
  {Gu},\ and\ \citenamefont {Duan}}]{23}%
  \BibitemOpen
  \bibfield  {author} {\bibinfo {author} {\bibfnamefont {C.}~\bibnamefont
  {Si}}, \bibinfo {author} {\bibfnamefont {J.}~\bibnamefont {Liu}}, \bibinfo
  {author} {\bibfnamefont {Y.}~\bibnamefont {Xu}}, \bibinfo {author}
  {\bibfnamefont {J.}~\bibnamefont {Wu}}, \bibinfo {author} {\bibfnamefont
  {B.-L.}\ \bibnamefont {Gu}}, \ and\ \bibinfo {author} {\bibfnamefont
  {W.}~\bibnamefont {Duan}},\ }\href@noop {} {\bibfield  {journal} {\bibinfo
  {journal} {Physical Review B}\ }\textbf {\bibinfo {volume} {89}},\ \bibinfo
  {pages} {115429} (\bibinfo {year} {2014})}\BibitemShut {NoStop}%
\bibitem [{\citenamefont {Weng}\ \emph {et~al.}(2015)\citenamefont {Weng},
  \citenamefont {Ranjbar}, \citenamefont {Liang}, \citenamefont {Song},
  \citenamefont {Khazaei}, \citenamefont {Yunoki}, \citenamefont {Arai},
  \citenamefont {Kawazoe}, \citenamefont {Fang},\ and\ \citenamefont
  {Dai}}]{24}%
  \BibitemOpen
  \bibfield  {author} {\bibinfo {author} {\bibfnamefont {H.}~\bibnamefont
  {Weng}}, \bibinfo {author} {\bibfnamefont {A.}~\bibnamefont {Ranjbar}},
  \bibinfo {author} {\bibfnamefont {Y.}~\bibnamefont {Liang}}, \bibinfo
  {author} {\bibfnamefont {Z.}~\bibnamefont {Song}}, \bibinfo {author}
  {\bibfnamefont {M.}~\bibnamefont {Khazaei}}, \bibinfo {author} {\bibfnamefont
  {S.}~\bibnamefont {Yunoki}}, \bibinfo {author} {\bibfnamefont
  {M.}~\bibnamefont {Arai}}, \bibinfo {author} {\bibfnamefont {Y.}~\bibnamefont
  {Kawazoe}}, \bibinfo {author} {\bibfnamefont {Z.}~\bibnamefont {Fang}}, \
  and\ \bibinfo {author} {\bibfnamefont {X.}~\bibnamefont {Dai}},\ }\href@noop
  {} {\bibfield  {journal} {\bibinfo  {journal} {Physical Review B}\ }\textbf
  {\bibinfo {volume} {92}},\ \bibinfo {pages} {075436} (\bibinfo {year}
  {2015})}\BibitemShut {NoStop}%
\bibitem [{\citenamefont {Wang}\ \emph {et~al.}(2017)\citenamefont {Wang},
  \citenamefont {Ji}, \citenamefont {Zhang}, \citenamefont {Li}, \citenamefont
  {Zhang}, \citenamefont {Wang}, \citenamefont {Li},\ and\ \citenamefont
  {Yan}}]{25}%
  \BibitemOpen
  \bibfield  {author} {\bibinfo {author} {\bibfnamefont {Y.-p.}\ \bibnamefont
  {Wang}}, \bibinfo {author} {\bibfnamefont {W.-x.}\ \bibnamefont {Ji}},
  \bibinfo {author} {\bibfnamefont {C.-w.}\ \bibnamefont {Zhang}}, \bibinfo
  {author} {\bibfnamefont {P.}~\bibnamefont {Li}}, \bibinfo {author}
  {\bibfnamefont {S.-f.}\ \bibnamefont {Zhang}}, \bibinfo {author}
  {\bibfnamefont {P.-j.}\ \bibnamefont {Wang}}, \bibinfo {author}
  {\bibfnamefont {S.-s.}\ \bibnamefont {Li}}, \ and\ \bibinfo {author}
  {\bibfnamefont {S.-s.}\ \bibnamefont {Yan}},\ }\href@noop {} {\bibfield
  {journal} {\bibinfo  {journal} {Applied Physics Letters}\ }\textbf {\bibinfo
  {volume} {110}},\ \bibinfo {pages} {213101} (\bibinfo {year}
  {2017})}\BibitemShut {NoStop}%
\bibitem [{\citenamefont {Liang}\ \emph {et~al.}(2017)\citenamefont {Liang},
  \citenamefont {Khazaei}, \citenamefont {Ranjbar}, \citenamefont {Arai},
  \citenamefont {Yunoki}, \citenamefont {Kawazoe}, \citenamefont {Weng},\ and\
  \citenamefont {Fang}}]{26}%
  \BibitemOpen
  \bibfield  {author} {\bibinfo {author} {\bibfnamefont {Y.}~\bibnamefont
  {Liang}}, \bibinfo {author} {\bibfnamefont {M.}~\bibnamefont {Khazaei}},
  \bibinfo {author} {\bibfnamefont {A.}~\bibnamefont {Ranjbar}}, \bibinfo
  {author} {\bibfnamefont {M.}~\bibnamefont {Arai}}, \bibinfo {author}
  {\bibfnamefont {S.}~\bibnamefont {Yunoki}}, \bibinfo {author} {\bibfnamefont
  {Y.}~\bibnamefont {Kawazoe}}, \bibinfo {author} {\bibfnamefont
  {H.}~\bibnamefont {Weng}}, \ and\ \bibinfo {author} {\bibfnamefont
  {Z.}~\bibnamefont {Fang}},\ }\href@noop {} {\bibfield  {journal} {\bibinfo
  {journal} {Physical Review B}\ }\textbf {\bibinfo {volume} {96}},\ \bibinfo
  {pages} {195414} (\bibinfo {year} {2017})}\BibitemShut {NoStop}%
\bibitem [{\citenamefont {Wu}, \citenamefont {Gu},\ and\ \citenamefont
  {Li}(2019)}]{27}%
  \BibitemOpen
  \bibfield  {author} {\bibinfo {author} {\bibfnamefont {L.}~\bibnamefont
  {Wu}}, \bibinfo {author} {\bibfnamefont {K.}~\bibnamefont {Gu}}, \ and\
  \bibinfo {author} {\bibfnamefont {Q.}~\bibnamefont {Li}},\ }\href@noop {}
  {\bibfield  {journal} {\bibinfo  {journal} {Applied Surface Science}\
  }\textbf {\bibinfo {volume} {484}},\ \bibinfo {pages} {1208--1213} (\bibinfo
  {year} {2019})}\BibitemShut {NoStop}%
\bibitem [{\citenamefont {Hu}, \citenamefont {Wang},\ and\ \citenamefont
  {Li}(2020)}]{28}%
  \BibitemOpen
  \bibfield  {author} {\bibinfo {author} {\bibfnamefont {X.-k.}\ \bibnamefont
  {Hu}}, \bibinfo {author} {\bibfnamefont {Y.}~\bibnamefont {Wang}}, \ and\
  \bibinfo {author} {\bibfnamefont {P.}~\bibnamefont {Li}},\ }\href@noop {}
  {\bibfield  {journal} {\bibinfo  {journal} {Chemical Physics Letters}\
  }\textbf {\bibinfo {volume} {740}},\ \bibinfo {pages} {137064} (\bibinfo
  {year} {2020})}\BibitemShut {NoStop}%
\bibitem [{\citenamefont {Huang}\ \emph {et~al.}(2020)\citenamefont {Huang},
  \citenamefont {Xu}, \citenamefont {Macam}, \citenamefont {Hsu},\ and\
  \citenamefont {Chuang}}]{29}%
  \BibitemOpen
  \bibfield  {author} {\bibinfo {author} {\bibfnamefont {Z.-Q.}\ \bibnamefont
  {Huang}}, \bibinfo {author} {\bibfnamefont {M.-L.}\ \bibnamefont {Xu}},
  \bibinfo {author} {\bibfnamefont {G.}~\bibnamefont {Macam}}, \bibinfo
  {author} {\bibfnamefont {C.-H.}\ \bibnamefont {Hsu}}, \ and\ \bibinfo
  {author} {\bibfnamefont {F.-C.}\ \bibnamefont {Chuang}},\ }\href@noop {}
  {\bibfield  {journal} {\bibinfo  {journal} {Physical Review B}\ }\textbf
  {\bibinfo {volume} {102}},\ \bibinfo {pages} {075306} (\bibinfo {year}
  {2020})}\BibitemShut {NoStop}%
\bibitem [{\citenamefont {Liu}\ \emph {et~al.}(2021)\citenamefont {Liu},
  \citenamefont {Gong}, \citenamefont {Li}, \citenamefont {Zhang},
  \citenamefont {He},\ and\ \citenamefont {Cao}}]{30}%
  \BibitemOpen
  \bibfield  {author} {\bibinfo {author} {\bibfnamefont {M.-Y.}\ \bibnamefont
  {Liu}}, \bibinfo {author} {\bibfnamefont {L.}~\bibnamefont {Gong}}, \bibinfo
  {author} {\bibfnamefont {W.-Z.}\ \bibnamefont {Li}}, \bibinfo {author}
  {\bibfnamefont {M.-L.}\ \bibnamefont {Zhang}}, \bibinfo {author}
  {\bibfnamefont {Y.}~\bibnamefont {He}}, \ and\ \bibinfo {author}
  {\bibfnamefont {C.}~\bibnamefont {Cao}},\ }\href@noop {} {\bibfield
  {journal} {\bibinfo  {journal} {Applied Surface Science}\ }\textbf {\bibinfo
  {volume} {540}},\ \bibinfo {pages} {148268} (\bibinfo {year}
  {2021})}\BibitemShut {NoStop}%
\bibitem [{\citenamefont {Miao}\ \emph {et~al.}(2015)\citenamefont {Miao},
  \citenamefont {Yao}, \citenamefont {Ming}, \citenamefont {Zhu}, \citenamefont
  {Han}, \citenamefont {Wang}, \citenamefont {Guan}, \citenamefont {Gao},
  \citenamefont {Liu}, \citenamefont {Liu} \emph {et~al.}}]{31}%
  \BibitemOpen
  \bibfield  {author} {\bibinfo {author} {\bibfnamefont {L.}~\bibnamefont
  {Miao}}, \bibinfo {author} {\bibfnamefont {M.-Y.}\ \bibnamefont {Yao}},
  \bibinfo {author} {\bibfnamefont {W.}~\bibnamefont {Ming}}, \bibinfo {author}
  {\bibfnamefont {F.}~\bibnamefont {Zhu}}, \bibinfo {author} {\bibfnamefont
  {C.}~\bibnamefont {Han}}, \bibinfo {author} {\bibfnamefont {Z.}~\bibnamefont
  {Wang}}, \bibinfo {author} {\bibfnamefont {D.}~\bibnamefont {Guan}}, \bibinfo
  {author} {\bibfnamefont {C.}~\bibnamefont {Gao}}, \bibinfo {author}
  {\bibfnamefont {C.}~\bibnamefont {Liu}}, \bibinfo {author} {\bibfnamefont
  {F.}~\bibnamefont {Liu}},  \emph {et~al.},\ }\href@noop {} {\bibfield
  {journal} {\bibinfo  {journal} {Physical Review B}\ }\textbf {\bibinfo
  {volume} {91}},\ \bibinfo {pages} {205414} (\bibinfo {year}
  {2015})}\BibitemShut {NoStop}%
\bibitem [{\citenamefont {Huang}\ \emph {et~al.}(2017)\citenamefont {Huang},
  \citenamefont {Guan}, \citenamefont {Lin}, \citenamefont {Liu}, \citenamefont
  {Xing}, \citenamefont {Wang},\ and\ \citenamefont {Guo}}]{32}%
  \BibitemOpen
  \bibfield  {author} {\bibinfo {author} {\bibfnamefont {X.}~\bibnamefont
  {Huang}}, \bibinfo {author} {\bibfnamefont {J.}~\bibnamefont {Guan}},
  \bibinfo {author} {\bibfnamefont {Z.}~\bibnamefont {Lin}}, \bibinfo {author}
  {\bibfnamefont {B.}~\bibnamefont {Liu}}, \bibinfo {author} {\bibfnamefont
  {S.}~\bibnamefont {Xing}}, \bibinfo {author} {\bibfnamefont {W.}~\bibnamefont
  {Wang}}, \ and\ \bibinfo {author} {\bibfnamefont {J.}~\bibnamefont {Guo}},\
  }\href@noop {} {\bibfield  {journal} {\bibinfo  {journal} {Nano letters}\
  }\textbf {\bibinfo {volume} {17}},\ \bibinfo {pages} {4619--4623} (\bibinfo
  {year} {2017})}\BibitemShut {NoStop}%
\bibitem [{\citenamefont {Zhu}\ \emph {et~al.}(2017)\citenamefont {Zhu},
  \citenamefont {Cai}, \citenamefont {Yi}, \citenamefont {Chen}, \citenamefont
  {Dai}, \citenamefont {Niu}, \citenamefont {Guo}, \citenamefont {Xie},
  \citenamefont {Liu}, \citenamefont {Cho} \emph {et~al.}}]{33}%
  \BibitemOpen
  \bibfield  {author} {\bibinfo {author} {\bibfnamefont {Z.}~\bibnamefont
  {Zhu}}, \bibinfo {author} {\bibfnamefont {X.}~\bibnamefont {Cai}}, \bibinfo
  {author} {\bibfnamefont {S.}~\bibnamefont {Yi}}, \bibinfo {author}
  {\bibfnamefont {J.}~\bibnamefont {Chen}}, \bibinfo {author} {\bibfnamefont
  {Y.}~\bibnamefont {Dai}}, \bibinfo {author} {\bibfnamefont {C.}~\bibnamefont
  {Niu}}, \bibinfo {author} {\bibfnamefont {Z.}~\bibnamefont {Guo}}, \bibinfo
  {author} {\bibfnamefont {M.}~\bibnamefont {Xie}}, \bibinfo {author}
  {\bibfnamefont {F.}~\bibnamefont {Liu}}, \bibinfo {author} {\bibfnamefont
  {J.-H.}\ \bibnamefont {Cho}},  \emph {et~al.},\ }\href@noop {} {\bibfield
  {journal} {\bibinfo  {journal} {Physical review letters}\ }\textbf {\bibinfo
  {volume} {119}},\ \bibinfo {pages} {106101} (\bibinfo {year}
  {2017})}\BibitemShut {NoStop}%
\bibitem [{\citenamefont {Qin}\ \emph {et~al.}(2017)\citenamefont {Qin},
  \citenamefont {Qiu}, \citenamefont {Jian}, \citenamefont {Zhou},
  \citenamefont {Yang}, \citenamefont {Charnas}, \citenamefont {Zemlyanov},
  \citenamefont {Xu}, \citenamefont {Xu}, \citenamefont {Wu} \emph
  {et~al.}}]{34}%
  \BibitemOpen
  \bibfield  {author} {\bibinfo {author} {\bibfnamefont {J.}~\bibnamefont
  {Qin}}, \bibinfo {author} {\bibfnamefont {G.}~\bibnamefont {Qiu}}, \bibinfo
  {author} {\bibfnamefont {J.}~\bibnamefont {Jian}}, \bibinfo {author}
  {\bibfnamefont {H.}~\bibnamefont {Zhou}}, \bibinfo {author} {\bibfnamefont
  {L.}~\bibnamefont {Yang}}, \bibinfo {author} {\bibfnamefont {A.}~\bibnamefont
  {Charnas}}, \bibinfo {author} {\bibfnamefont {D.~Y.}\ \bibnamefont
  {Zemlyanov}}, \bibinfo {author} {\bibfnamefont {C.-Y.}\ \bibnamefont {Xu}},
  \bibinfo {author} {\bibfnamefont {X.}~\bibnamefont {Xu}}, \bibinfo {author}
  {\bibfnamefont {W.}~\bibnamefont {Wu}},  \emph {et~al.},\ }\href@noop {}
  {\bibfield  {journal} {\bibinfo  {journal} {ACS nano}\ }\textbf {\bibinfo
  {volume} {11}},\ \bibinfo {pages} {10222--10229} (\bibinfo {year}
  {2017})}\BibitemShut {NoStop}%
\bibitem [{\citenamefont {Zhao}\ and\ \citenamefont {Wang}(2020)}]{35}%
  \BibitemOpen
  \bibfield  {author} {\bibinfo {author} {\bibfnamefont {A.}~\bibnamefont
  {Zhao}}\ and\ \bibinfo {author} {\bibfnamefont {B.}~\bibnamefont {Wang}},\
  }\href@noop {} {\bibfield  {journal} {\bibinfo  {journal} {APL Materials}\
  }\textbf {\bibinfo {volume} {8}},\ \bibinfo {pages} {030701} (\bibinfo {year}
  {2020})}\BibitemShut {NoStop}%
\bibitem [{\citenamefont {Bouaziz}\ \emph {et~al.}(2020)\citenamefont
  {Bouaziz}, \citenamefont {Zhang}, \citenamefont {Tong}, \citenamefont
  {Oughaddou}, \citenamefont {Enriquez}, \citenamefont {Mlika}, \citenamefont
  {Korri-Youssoufi}, \citenamefont {Chen}, \citenamefont {Xiong}, \citenamefont
  {Cheng} \emph {et~al.}}]{36}%
  \BibitemOpen
  \bibfield  {author} {\bibinfo {author} {\bibfnamefont {M.}~\bibnamefont
  {Bouaziz}}, \bibinfo {author} {\bibfnamefont {W.}~\bibnamefont {Zhang}},
  \bibinfo {author} {\bibfnamefont {Y.}~\bibnamefont {Tong}}, \bibinfo {author}
  {\bibfnamefont {H.}~\bibnamefont {Oughaddou}}, \bibinfo {author}
  {\bibfnamefont {H.}~\bibnamefont {Enriquez}}, \bibinfo {author}
  {\bibfnamefont {R.}~\bibnamefont {Mlika}}, \bibinfo {author} {\bibfnamefont
  {H.}~\bibnamefont {Korri-Youssoufi}}, \bibinfo {author} {\bibfnamefont
  {Z.}~\bibnamefont {Chen}}, \bibinfo {author} {\bibfnamefont {H.}~\bibnamefont
  {Xiong}}, \bibinfo {author} {\bibfnamefont {Y.}~\bibnamefont {Cheng}},  \emph
  {et~al.},\ }\href@noop {} {\bibfield  {journal} {\bibinfo  {journal} {2D
  Materials}\ }\textbf {\bibinfo {volume} {8}},\ \bibinfo {pages} {015029}
  (\bibinfo {year} {2020})}\BibitemShut {NoStop}%
\bibitem [{\citenamefont {Giannozzi}\ \emph {et~al.}(2009)\citenamefont
  {Giannozzi}, \citenamefont {Baroni}, \citenamefont {Bonini}, \citenamefont
  {Calandra}, \citenamefont {Car}, \citenamefont {Cavazzoni}, \citenamefont
  {Ceresoli}, \citenamefont {Chiarotti}, \citenamefont {Cococcioni},
  \citenamefont {Dabo} \emph {et~al.}}]{37}%
  \BibitemOpen
  \bibfield  {author} {\bibinfo {author} {\bibfnamefont {P.}~\bibnamefont
  {Giannozzi}}, \bibinfo {author} {\bibfnamefont {S.}~\bibnamefont {Baroni}},
  \bibinfo {author} {\bibfnamefont {N.}~\bibnamefont {Bonini}}, \bibinfo
  {author} {\bibfnamefont {M.}~\bibnamefont {Calandra}}, \bibinfo {author}
  {\bibfnamefont {R.}~\bibnamefont {Car}}, \bibinfo {author} {\bibfnamefont
  {C.}~\bibnamefont {Cavazzoni}}, \bibinfo {author} {\bibfnamefont
  {D.}~\bibnamefont {Ceresoli}}, \bibinfo {author} {\bibfnamefont {G.~L.}\
  \bibnamefont {Chiarotti}}, \bibinfo {author} {\bibfnamefont {M.}~\bibnamefont
  {Cococcioni}}, \bibinfo {author} {\bibfnamefont {I.}~\bibnamefont {Dabo}},
  \emph {et~al.},\ }\href@noop {} {\bibfield  {journal} {\bibinfo  {journal}
  {Journal of physics: Condensed matter}\ }\textbf {\bibinfo {volume} {21}},\
  \bibinfo {pages} {395502} (\bibinfo {year} {2009})}\BibitemShut {NoStop}%
\bibitem [{\citenamefont {Perdew}, \citenamefont {Burke},\ and\ \citenamefont
  {Ernzerhof}(1996)}]{38}%
  \BibitemOpen
  \bibfield  {author} {\bibinfo {author} {\bibfnamefont {J.~P.}\ \bibnamefont
  {Perdew}}, \bibinfo {author} {\bibfnamefont {K.}~\bibnamefont {Burke}}, \
  and\ \bibinfo {author} {\bibfnamefont {M.}~\bibnamefont {Ernzerhof}},\
  }\href@noop {} {\bibfield  {journal} {\bibinfo  {journal} {Physical review
  letters}\ }\textbf {\bibinfo {volume} {77}},\ \bibinfo {pages} {3865}
  (\bibinfo {year} {1996})}\BibitemShut {NoStop}%
\bibitem [{\citenamefont {Bl{\"o}chl}(1994)}]{39}%
  \BibitemOpen
  \bibfield  {author} {\bibinfo {author} {\bibfnamefont {P.~E.}\ \bibnamefont
  {Bl{\"o}chl}},\ }\href@noop {} {\bibfield  {journal} {\bibinfo  {journal}
  {Physical review B}\ }\textbf {\bibinfo {volume} {50}},\ \bibinfo {pages}
  {17953} (\bibinfo {year} {1994})}\BibitemShut {NoStop}%
\bibitem [{\citenamefont {Monkhorst}\ and\ \citenamefont {Pack}(1976)}]{40}%
  \BibitemOpen
  \bibfield  {author} {\bibinfo {author} {\bibfnamefont {H.~J.}\ \bibnamefont
  {Monkhorst}}\ and\ \bibinfo {author} {\bibfnamefont {J.~D.}\ \bibnamefont
  {Pack}},\ }\href@noop {} {\bibfield  {journal} {\bibinfo  {journal} {Physical
  review B}\ }\textbf {\bibinfo {volume} {13}},\ \bibinfo {pages} {5188}
  (\bibinfo {year} {1976})}\BibitemShut {NoStop}%
\bibitem [{\citenamefont {Baroni}\ \emph {et~al.}(2001)\citenamefont {Baroni},
  \citenamefont {De~Gironcoli}, \citenamefont {Dal~Corso},\ and\ \citenamefont
  {Giannozzi}}]{41}%
  \BibitemOpen
  \bibfield  {author} {\bibinfo {author} {\bibfnamefont {S.}~\bibnamefont
  {Baroni}}, \bibinfo {author} {\bibfnamefont {S.}~\bibnamefont
  {De~Gironcoli}}, \bibinfo {author} {\bibfnamefont {A.}~\bibnamefont
  {Dal~Corso}}, \ and\ \bibinfo {author} {\bibfnamefont {P.}~\bibnamefont
  {Giannozzi}},\ }\href@noop {} {\bibfield  {journal} {\bibinfo  {journal}
  {Reviews of modern Physics}\ }\textbf {\bibinfo {volume} {73}},\ \bibinfo
  {pages} {515} (\bibinfo {year} {2001})}\BibitemShut {NoStop}%
\bibitem [{\citenamefont {Mostofi}\ \emph {et~al.}(2014)\citenamefont
  {Mostofi}, \citenamefont {Yates}, \citenamefont {Pizzi}, \citenamefont {Lee},
  \citenamefont {Souza}, \citenamefont {Vanderbilt},\ and\ \citenamefont
  {Marzari}}]{42}%
  \BibitemOpen
  \bibfield  {author} {\bibinfo {author} {\bibfnamefont {A.~A.}\ \bibnamefont
  {Mostofi}}, \bibinfo {author} {\bibfnamefont {J.~R.}\ \bibnamefont {Yates}},
  \bibinfo {author} {\bibfnamefont {G.}~\bibnamefont {Pizzi}}, \bibinfo
  {author} {\bibfnamefont {Y.-S.}\ \bibnamefont {Lee}}, \bibinfo {author}
  {\bibfnamefont {I.}~\bibnamefont {Souza}}, \bibinfo {author} {\bibfnamefont
  {D.}~\bibnamefont {Vanderbilt}}, \ and\ \bibinfo {author} {\bibfnamefont
  {N.}~\bibnamefont {Marzari}},\ }\href@noop {} {\bibfield  {journal} {\bibinfo
   {journal} {Computer Physics Communications}\ }\textbf {\bibinfo {volume}
  {185}},\ \bibinfo {pages} {2309--2310} (\bibinfo {year} {2014})}\BibitemShut
  {NoStop}%
\bibitem [{\citenamefont {Wu}\ \emph {et~al.}(2018)\citenamefont {Wu},
  \citenamefont {Zhang}, \citenamefont {Song}, \citenamefont {Troyer},\ and\
  \citenamefont {Soluyanov}}]{43}%
  \BibitemOpen
  \bibfield  {author} {\bibinfo {author} {\bibfnamefont {Q.}~\bibnamefont
  {Wu}}, \bibinfo {author} {\bibfnamefont {S.}~\bibnamefont {Zhang}}, \bibinfo
  {author} {\bibfnamefont {H.-F.}\ \bibnamefont {Song}}, \bibinfo {author}
  {\bibfnamefont {M.}~\bibnamefont {Troyer}}, \ and\ \bibinfo {author}
  {\bibfnamefont {A.~A.}\ \bibnamefont {Soluyanov}},\ }\href@noop {} {\bibfield
   {journal} {\bibinfo  {journal} {Computer Physics Communications}\ }\textbf
  {\bibinfo {volume} {224}},\ \bibinfo {pages} {405--416} (\bibinfo {year}
  {2018})}\BibitemShut {NoStop}%
\bibitem [{\citenamefont {Qiao}\ \emph {et~al.}(2018)\citenamefont {Qiao},
  \citenamefont {Zhou}, \citenamefont {Yuan},\ and\ \citenamefont {Zhao}}]{44}%
  \BibitemOpen
  \bibfield  {author} {\bibinfo {author} {\bibfnamefont {J.}~\bibnamefont
  {Qiao}}, \bibinfo {author} {\bibfnamefont {J.}~\bibnamefont {Zhou}}, \bibinfo
  {author} {\bibfnamefont {Z.}~\bibnamefont {Yuan}}, \ and\ \bibinfo {author}
  {\bibfnamefont {W.}~\bibnamefont {Zhao}},\ }\href@noop {} {\bibfield
  {journal} {\bibinfo  {journal} {Physical Review B}\ }\textbf {\bibinfo
  {volume} {98}},\ \bibinfo {pages} {214402} (\bibinfo {year}
  {2018})}\BibitemShut {NoStop}%
\bibitem [{\citenamefont {Guo}\ \emph {et~al.}(2008)\citenamefont {Guo},
  \citenamefont {Murakami}, \citenamefont {Chen},\ and\ \citenamefont
  {Nagaosa}}]{45}%
  \BibitemOpen
  \bibfield  {author} {\bibinfo {author} {\bibfnamefont {G.-Y.}\ \bibnamefont
  {Guo}}, \bibinfo {author} {\bibfnamefont {S.}~\bibnamefont {Murakami}},
  \bibinfo {author} {\bibfnamefont {T.-W.}\ \bibnamefont {Chen}}, \ and\
  \bibinfo {author} {\bibfnamefont {N.}~\bibnamefont {Nagaosa}},\ }\href@noop
  {} {\bibfield  {journal} {\bibinfo  {journal} {Physical review letters}\
  }\textbf {\bibinfo {volume} {100}},\ \bibinfo {pages} {096401} (\bibinfo
  {year} {2008})}\BibitemShut {NoStop}%
\bibitem [{\citenamefont {Sui}\ \emph {et~al.}(2017)\citenamefont {Sui},
  \citenamefont {Wang}, \citenamefont {Kim}, \citenamefont {Wang},
  \citenamefont {Rhim}, \citenamefont {Duan},\ and\ \citenamefont
  {Kioussis}}]{46}%
  \BibitemOpen
  \bibfield  {author} {\bibinfo {author} {\bibfnamefont {X.}~\bibnamefont
  {Sui}}, \bibinfo {author} {\bibfnamefont {C.}~\bibnamefont {Wang}}, \bibinfo
  {author} {\bibfnamefont {J.}~\bibnamefont {Kim}}, \bibinfo {author}
  {\bibfnamefont {J.}~\bibnamefont {Wang}}, \bibinfo {author} {\bibfnamefont
  {S.}~\bibnamefont {Rhim}}, \bibinfo {author} {\bibfnamefont {W.}~\bibnamefont
  {Duan}}, \ and\ \bibinfo {author} {\bibfnamefont {N.}~\bibnamefont
  {Kioussis}},\ }\href@noop {} {\bibfield  {journal} {\bibinfo  {journal}
  {Physical Review B}\ }\textbf {\bibinfo {volume} {96}},\ \bibinfo {pages}
  {241105} (\bibinfo {year} {2017})}\BibitemShut {NoStop}%
\bibitem [{\citenamefont {Wang}\ \emph
  {et~al.}(2016{\natexlab{c}})\citenamefont {Wang}, \citenamefont {Chen},
  \citenamefont {Shi}, \citenamefont {Wang}, \citenamefont {Cui}, \citenamefont
  {Zhang},\ and\ \citenamefont {Chen}}]{47}%
  \BibitemOpen
  \bibfield  {author} {\bibinfo {author} {\bibfnamefont {D.}~\bibnamefont
  {Wang}}, \bibinfo {author} {\bibfnamefont {L.}~\bibnamefont {Chen}}, \bibinfo
  {author} {\bibfnamefont {C.}~\bibnamefont {Shi}}, \bibinfo {author}
  {\bibfnamefont {X.}~\bibnamefont {Wang}}, \bibinfo {author} {\bibfnamefont
  {G.}~\bibnamefont {Cui}}, \bibinfo {author} {\bibfnamefont {P.}~\bibnamefont
  {Zhang}}, \ and\ \bibinfo {author} {\bibfnamefont {Y.}~\bibnamefont {Chen}},\
  }\href@noop {} {\bibfield  {journal} {\bibinfo  {journal} {Scientific
  reports}\ }\textbf {\bibinfo {volume} {6}},\ \bibinfo {pages} {1--7}
  (\bibinfo {year} {2016}{\natexlab{c}})}\BibitemShut {NoStop}%
\bibitem [{\citenamefont {Farzaneh}\ and\ \citenamefont {Rakheja}(2020)}]{48}%
  \BibitemOpen
  \bibfield  {author} {\bibinfo {author} {\bibfnamefont {S.}~\bibnamefont
  {Farzaneh}}\ and\ \bibinfo {author} {\bibfnamefont {S.}~\bibnamefont
  {Rakheja}},\ }\href@noop {} {\bibfield  {journal} {\bibinfo  {journal} {arXiv
  preprint arXiv:2009.13753}\ } (\bibinfo {year} {2020})}\BibitemShut {NoStop}%
\bibitem [{\citenamefont {Matusalem}\ \emph {et~al.}(2019)\citenamefont
  {Matusalem}, \citenamefont {Marques}, \citenamefont {Teles}, \citenamefont
  {Matthes}, \citenamefont {Furthm{\"u}ller},\ and\ \citenamefont
  {Bechstedt}}]{49}%
  \BibitemOpen
  \bibfield  {author} {\bibinfo {author} {\bibfnamefont {F.}~\bibnamefont
  {Matusalem}}, \bibinfo {author} {\bibfnamefont {M.}~\bibnamefont {Marques}},
  \bibinfo {author} {\bibfnamefont {L.~K.}\ \bibnamefont {Teles}}, \bibinfo
  {author} {\bibfnamefont {L.}~\bibnamefont {Matthes}}, \bibinfo {author}
  {\bibfnamefont {J.}~\bibnamefont {Furthm{\"u}ller}}, \ and\ \bibinfo {author}
  {\bibfnamefont {F.}~\bibnamefont {Bechstedt}},\ }\href@noop {} {\bibfield
  {journal} {\bibinfo  {journal} {Physical Review B}\ }\textbf {\bibinfo
  {volume} {100}},\ \bibinfo {pages} {245430} (\bibinfo {year}
  {2019})}\BibitemShut {NoStop}%
\bibitem [{\citenamefont {Zhou}\ \emph {et~al.}(2018)\citenamefont {Zhou},
  \citenamefont {Zhang}, \citenamefont {Jiang}, \citenamefont
  {{\v{Z}}uti{\'c}},\ and\ \citenamefont {Yang}}]{50}%
  \BibitemOpen
  \bibfield  {author} {\bibinfo {author} {\bibfnamefont {T.}~\bibnamefont
  {Zhou}}, \bibinfo {author} {\bibfnamefont {J.}~\bibnamefont {Zhang}},
  \bibinfo {author} {\bibfnamefont {H.}~\bibnamefont {Jiang}}, \bibinfo
  {author} {\bibfnamefont {I.}~\bibnamefont {{\v{Z}}uti{\'c}}}, \ and\ \bibinfo
  {author} {\bibfnamefont {Z.}~\bibnamefont {Yang}},\ }\href@noop {} {\bibfield
   {journal} {\bibinfo  {journal} {npj Quantum Materials}\ }\textbf {\bibinfo
  {volume} {3}},\ \bibinfo {pages} {1--7} (\bibinfo {year} {2018})}\BibitemShut
  {NoStop}%
\bibitem [{\citenamefont {Grimme}\ \emph {et~al.}(2010)\citenamefont {Grimme},
  \citenamefont {Antony}, \citenamefont {Ehrlich},\ and\ \citenamefont
  {Krieg}}]{51}%
  \BibitemOpen
  \bibfield  {author} {\bibinfo {author} {\bibfnamefont {S.}~\bibnamefont
  {Grimme}}, \bibinfo {author} {\bibfnamefont {J.}~\bibnamefont {Antony}},
  \bibinfo {author} {\bibfnamefont {S.}~\bibnamefont {Ehrlich}}, \ and\
  \bibinfo {author} {\bibfnamefont {H.}~\bibnamefont {Krieg}},\ }\href@noop {}
  {\bibfield  {journal} {\bibinfo  {journal} {The Journal of chemical physics}\
  }\textbf {\bibinfo {volume} {132}},\ \bibinfo {pages} {154104} (\bibinfo
  {year} {2010})}\BibitemShut {NoStop}%
\end{thebibliography}%

\newpage
\section*{SUPPLEMENTARY MATERIAL}

\subsection*{Structure, phonon dispersion curves and electronic band structure:}

\begin{figure}[ht]
	\begin{center}
		\includegraphics[width=0.95\textwidth]{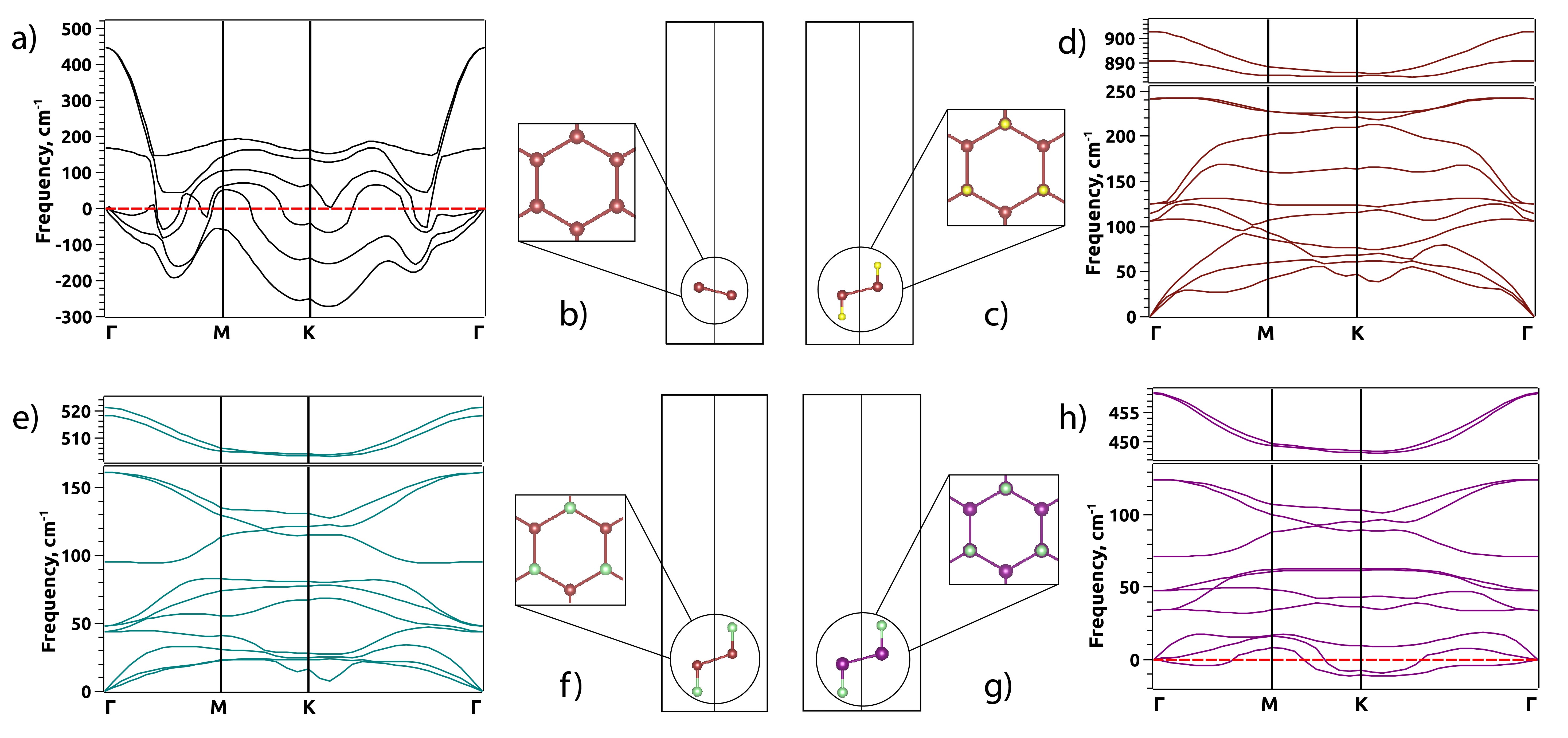}
		\caption{Honeycomb lattice structure alongside phonon dispersion curves of \textbf{(a, b)} Se, \textbf{(c, d)} SeO, \textbf{(e, f)} SeS and \textbf{(g, h)} TeS.}
		\label{sm1}
	\end{center}
\end{figure}

Figure \ref{sm1} presents the predicted honeycomb lattice structure and phonon dispersion curves of Se (Fig. \ref{sm1}(a, b)), SeO (Fig. \ref{sm1}(c, d)), SeS (Fig. \ref{sm1}(e, f)) and TeS Fig. (\ref{sm1}(g, h)) which clearly indicates that, SeO and SeS are dynamically stable (since imaginary modes are absent in the entire Brillouin Zone (BZ)) whereas Se and TeS are dynamically unstable which can be attributed to the soft modes evident from the imaginary modes in the BZ. Also, from the phonon dispersion curves in Fig. \ref{sm1}(d) and (e) we observe softening of the phonon modes along the M $\rightarrow$ K $\rightarrow$ $\Gamma$ high symmetry points in the BZ which indicates structural instability at higher temperatures.

\begin{table}[h!]
	\caption{\label{tab}
		Lattice constants (\textit{a}) and structural parameters of 2D systems of type A and AB where, A = Se, Te and B = O, S (with angles, buckling height (h), layer thickness (\textit{t}) and distance between A and B (\textit{d})).}
	\begin{ruledtabular}
		\begin{tabular}{ccccccc}
			AB (A = Se, Te; B = O, S) & \textit{a} (\AA) & $\angle$BAA & $\angle$AAA & \textit{h} (\AA) & \textit{t} (\AA) & \textit{d} (\AA)\\
			\hline
			Se & 4.36 & $ - $ & 114.29$^{\circ}$ & 0.63 & $ - $ & $ - $\\
			SeO & 4.79 & 103.87$^{\circ}$ & 114.44$^{\circ}$ & 0.68 & 4.00 & 1.66\\
			SeS & 4.80 & 108.12$^{\circ}$ & 110.79$^{\circ}$ & 0.91 & 5.04 & 2.06\\
			TeS & 5.36 & 104.68$^{\circ}$ & 113.80$^{\circ}$ & 0.81 & 5.28 & 2.23\\
		\end{tabular}
	\end{ruledtabular}
\end{table}

\begin{figure}[ht]
	\begin{center}
		\includegraphics[width=0.75\textwidth]{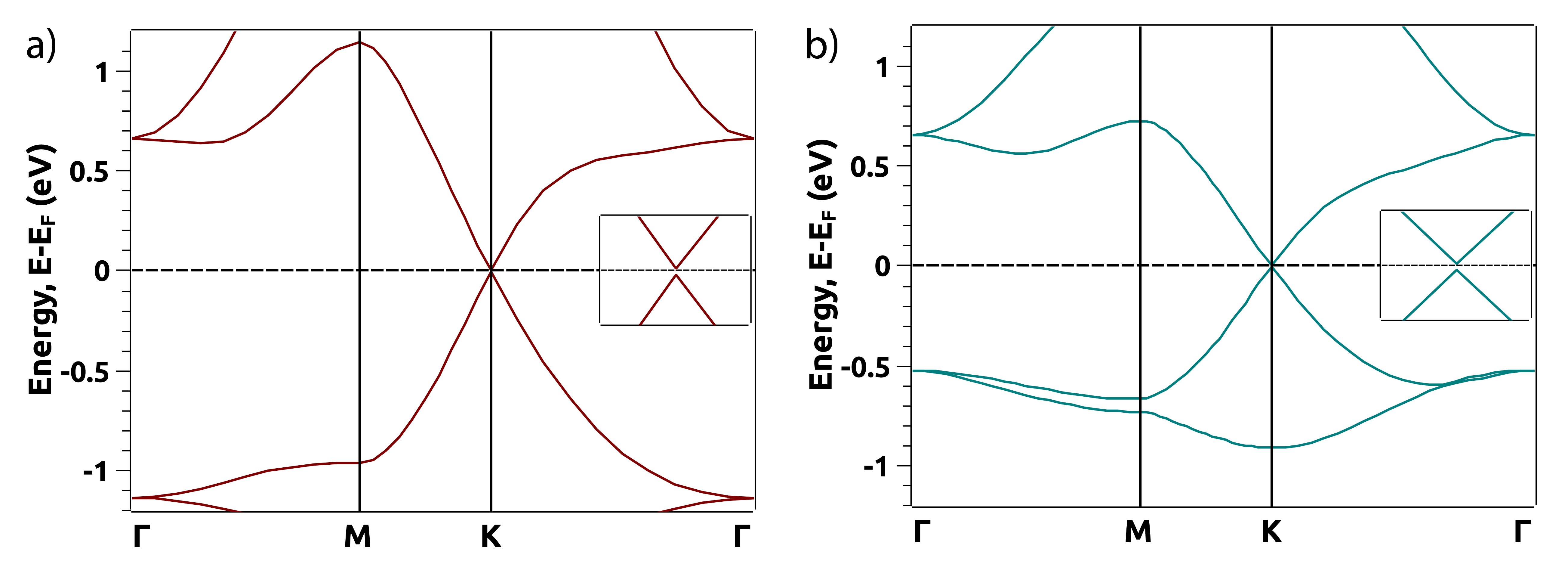}
		\caption{Electronic band structure of \textbf{(a)} SeO and \textbf{(b)} SeS with inset indicating the semi-conducting gap along the high symmetry point K in the BZ.}
		\label{sm2}
	\end{center}
\end{figure}

The optimized lattice constants and structural parameters of the proposed (stable and unstable) two-dimensional (2D) systems are tabulated in Table \ref{tab}. Since, Se and TeS are dynamically unstable, we present the electronic band structure of SeO and SeS only. These systems exhibit semi-conducting nature (as evident from the electronic band structure presented in Fig. \ref{sm2}) with gaps of the order of $\approx$ 10$^{-3}$ eV (SeO $\approx$ 1.8 meV, SeS $\approx$ 2.3 meV) which can be exploited to realise topologically non-trivial behaviour under the influence of modest tensile/compressive strain or external electric field or under influence of spin-orbit coupling effects.

\end{document}